\documentclass[showpacs,twocolumn,aps,superscriptaddress]{revtex4}
\usepackage{graphicx}
\usepackage{palatino}
\usepackage[latin1]{inputenc}
\usepackage{epsf}
\usepackage{dsfont}
\usepackage{amsmath,amssymb}
\usepackage{latexsym}
\usepackage{calc}
\usepackage{color}
\usepackage{shadow}
\usepackage{epsfig}

\newcommand{\ben}{\begin{equation}}
\newcommand{\een}{\end{equation}}
\newcommand{\bea}{\begin{eqnarray}}
\newcommand{\eea}{\end{eqnarray}}

\def\n{n}

\def\sss{\scriptscriptstyle\rm}

\def\s{_{\sss S}}

\def\xc{_{\sss XC}}

\def\Hxc{_{\sss HXC}}

\def\H{_{\sss H}}

\def\ext{_{\rm ext}}

\def\ee{_{\rm ee}}

\def\2c{_{2{\sss C}}}

\def\br{{\bf r}}

\def\bp{{\bf p}}
\def\bq{{\bf q}}
\def\by{{\bf y}}
\def\half{\frac{1}{2}}

\begin{document}
  \title{Perspectives on double-excitations in TDDFT}
  \author{Peter Elliott}
\affiliation{Department of Physics and Astronomy, Hunter College and the City University of New York, 695 Park Avenue, New York, New York 10065, USA}
\author{Sharma Goldson}
\affiliation{Department of Physics and Astronomy, Hunter College and the City University of New York, 695 Park Avenue, New York, New York 10065, USA}
\author{Chris Canahui}
\affiliation{Department of Physics and Astronomy, Hunter College and the City University of New York, 695 Park Avenue, New York, New York 10065, USA}
\author{Neepa T. Maitra}
\affiliation{Department of Physics and Astronomy, Hunter College and the City University of New York, 695 Park Avenue, New York, New York 10065, USA}

  \date{\today}

 \begin{abstract}
The adiabatic approximation in time-dependent density functional
theory (TDDFT) yields reliable excitation spectra with great
efficiency in many cases, but fundamentally fails for
states of double-excitation character. We discuss how double-excitations are at the root
of some of the most challenging problems for TDDFT today. We then present new results for (i) the calculation of autoionizing
resonances in the helium atom, (ii) understanding the nature of the
double excitations appearing in the quadratic response function, and
(iii) retrieving double-excitations through a real-time semiclassical
approach to correlation in a model quantum dot.
 \end{abstract}
 \maketitle 

\section{Introduction}
There is no question that time-dependent density functional theory (TDDFT) has greatly impacted calculations of
excitations and spectra of a wide range of systems, from atoms and
molecules, to biomolecules and solids~\cite{TDDFTbook,RG84}. Its successes
have encouraged bold and exciting applications to study systems as
complex as photosynthetic processes in biomolecules, coupled
electron-ion dynamics after photoexcitation, molecular
transport, e.g. Refs.~\cite{TTRF08,KCBC08,EFB07,SRVB09}.  The usual approximations used
for the exchange-correlation (xc) potential in TDDFT calculations
however perform poorly in a number of situations particularly relevant
for some of these applications, e.g. states of double-excitation
character~\cite{MZCB04}, long-range charge-transfer
excitations~\cite{DWH03,T03}, conical
intersections~\cite{LKQM06,TTRF08}, polarizabilities of long-chain
polymers~\cite{FBLB02}, optical response in
solids~\cite{ORR02,GORT07}. Improved functionals, modeled from
first-principles, have been developed, and are beginning to be used,
to treat some of these. In this paper, we focus on the problem of double-excitations in TDDFT, for which recently much progress has been made. 

The term ``double-excitation'' is a short-hand for ``state of
double-excitation character''. In a non-interacting picture, such as
Kohn-Sham (KS), a double-excitation means one in which two electrons
have been promoted out of orbitals occupied in the ground-state, to
two virtual orbitals, forming a ``doubly-excited'' Slater determinant.
The picture of placing electrons in single-particle orbitals however
does not apply to true interacting states. Instead one may expand any
interacting state as a linear combination of Slater determinants, say
eigenstates of some one-body Hamiltonian, such as the KS Hamiltonian,
and then a state of double-excitation character is one which has a
significant proportion of a doubly-excited Slater
determinant. Clearly, the exact value of this proportion depends on
the level of theory used for the ground-state reference e.g. it will
be different in Hartree-Fock than in TDDFT.

Given the important role of correlation in states of double-excitation
character, the question of how these appear in a single-particle-based
theory has both fundamental and practical interest. We shall begin by
reviewing the status of linear-response TDDFT in this regard: one must
go beyond the ubiquitous adiabatic approximation to capture these. We
conclude the introduction by discussing systems where
double-excitations are particularly important, and we shall see that
some of these cases are due to the peculiarity of the KS single-Slater
determinant in the ground-state (e.g. in certain long-range
charge-transfer states). Then in Section~\ref{sec:autoionizing} we
consider what happens when the double-excitation lies in the
continuum, and test a recently developed kernel approximation to
describe the resulting autoionizing resonance in the He atom. In
Section~\ref{sec:quadratic}, we study whether adiabatic kernels can be
redeemed for double-excitations by going to quadratic response theory.
Section~\ref{sec:frozengauss} turns to a new approach that was
recently proposed for general many-electron dynamics, that uses
semiclassical dynamics to evaluate electron correlation. Here we test it on a model system to see whether it captures double-excitations.

\subsection{Double excitations in linear-response TDDFT}
Excitation spectra can be obtained in two ways from TDDFT. In one, a
weak perturbation is applied to the KS system in its ground-state, and
the dynamics in real-time of, e.g. the dipole moment is Fourier
transformed to reveal peaks at the excitation frequencies of the
system, whose strengths indicate the oscillator strength. More often, a formulation directly in the frequency-domain is
used, in which two steps are performed: 
First,   the KS orbital energy differences 
between occupied ($i$) and unoccupied ($a$) orbitals, $\omega\s = \epsilon_a - \epsilon_i$ are computed. Second, these frequencies are corrected towards the true excitations through solution of a generalized eigenvalue problem~\cite{PGG96,C96}, utilizing the
Hartree-exchange-correlation kernel, $f\Hxc[n_0](\br,\br',\omega) = 1/\vert\br - \br'\vert + f\xc[n_0](\br,\br',\omega)$, a
functional of the ground-state density $n_0(\br)$.

Fundamentally, the origin of the linear response formalism is the
Dyson-like equation that links the density-density response function
of the non-interacting KS system, $\chi\s$, with that of the true
system, $\chi$:
\ben 
\chi(\omega) =\chi\s(\omega) +\chi\s(\omega)\star f\Hxc(\omega)\star\chi(\omega)
\label{eq:Dysonlin}
\een 
where we use the shorthand $\star$ to indicate the integral,  
$
\chi\s(\omega) \star f\Hxc(\omega) = \int d^3r_1 \ \chi\s(\br,\br_1,\omega)f\Hxc(\br_1,\br',\omega)
$
thinking of $\chi, \chi\s, f\Hxc$ etc as infinite-dimensional matrices in $\br,\br'$, each element of which is a function of $\omega$. 
The interacting density-density response function
$\chi[n_0](\br,\br',t-t') = \delta n(\br,t)/\delta
v(\br't')\vert_{n=n_0}$ measures the response in the density
$n(\br,t)$ to a perturbing external potential $v(\br,t)$. In the
frequency domain, 
\bea
\nonumber
\chi(\br,\br',\omega) &=&\sum_I \left(\frac{\langle \Psi_0\vert \hat{n}(\br) \vert\Psi_I\rangle\langle \Psi_I\vert \hat{n}(\br') \vert\Psi_0\rangle}{\omega - \omega_I + i0^+}\right. \\
&-& \left.\frac{\langle \Psi_0\vert \hat{n}(\br')\vert \Psi_I\rangle\langle \Psi_I\vert \hat{n}(\br)\vert \Psi_0\rangle}{\omega + \omega_I + i0^+}\right)
\label{eq:chi}
\eea
where $I$ labels the interacting excited-states, and $\omega_I = E_I - E_0$ is their frequency relative to the ground-state.
Similar expressions hold for the KS system, substituting the interacting wavefunctions above with KS single-Slater-determinants.

In almost all calculations, an adiabatic approximation is made for the
exchange-correlation (xc) kernel, i.e. one that is
frequency-independent, corresponding to an xc potential that depends
instantaneously on the density. But it is known that the exact kernel
is non-local in time, reflecting the xc potential's dependence on the
history of the density.
It is perhaps surprising that the adiabatic approximation works as well as it does, given that even weak excitations of a system lead it out of a ground-state (even if its density is that of a ground-state). 
One of the reasons for its success for
general excitation spectra is that the KS excitations themselves are
often themselves good zeroth-order approximations.  But this reason cannot
apply to double-excitations, since double-excitations are absent in
linear response of the KS system: to excite two electrons of a
non-interacting system two photons would be required, beyond linear
response. Only single-excitations of the KS system are available for
an adiabatic kernel to mix.  Indeed, if we consider the KS linear density
response function, the numerator of Eq.~\ref{eq:chi} with KS wavefunctions contains
$\langle\Phi_0\vert\hat{n}(\br)\vert\Phi_I\rangle$, which vanishes if the excited determinant $\Phi_I$ differs from the ground-state $\Phi_0$ by more than one orbital. The
one-body operator $\hat{n}(\br)$ cannot connect states that differ by
more than one orbital. The true response function, on the other hand,
retains poles at the true excitations which are mixtures of single,
double, and higher-electron-number excitations, as the numerator
$\langle\Psi_0\vert\hat{n}(\br)\vert\Psi_I\rangle$ remains finite due
to the mixed nature of both $\Psi_0$ and $\Psi_I$.  Within the
adiabatic approximation, $\chi$ therefore contains more poles than
$\chi\s$.

Ref.~\cite{TH00} pointed out the need to go beyond the adiabatic
approximation in capturing states of double-excitation
character. Ref.~\cite{MZCB04} derived a frequency-dependent kernel, motivated by first-principles, to be applied within the subspace of single KS excitations that mix strongly with the double-excitation of interest. For the case of one single-excitation, $q = i\to a$, coupled to one double-excitation:
\begin{equation}
2[q\vert f\xc(\omega)\vert q]=2[q\vert f\xc^{A}(\omega_q)\vert q]+\frac{|H_{qD}|^2}{\omega-(H_{DD}-H_{00})}.
\label{eq:genspa}
\end{equation}
to be applied within a dressed single-pole approximation (``DSPA''), 
\ben
\omega = \omega_q + 2[q\vert f\xc(\omega)\vert q]
\label{eq:spa}
\een
The Hamiltonian matrix elements in the dynamical correction
(second term of Eq.~(\ref{eq:genspa})) are those of the true interacting Hamiltonian, taken
between the single ($q$) and double ($D$) KS Slater determinants of
interest, as indicated, and $H_{00}$ is the expectation value of the
true Hamiltonian in the KS ground-state. This can be generalized to
cases where several single-excitations and double-excitations strongly mix, within a ``dressed Tamm-Dancoff'' scheme (see,
e.g.~\cite{CZMB04,MW09}). The kernel is to be applied as an {\it a
  posteriori} correction to a usual adiabatic calculation: first, one
scans over the KS orbital energies to see if the sum of two of their
frequencies lies near a single excitation frequency, and then applies
this kernel just to that pair.

Essentially the same formula results from derivations with different
starting points: in Ref.~\cite{C05}, it emerges as a polarization
propagator correction to adiabatic TDDFT in a superoperator formalism, made more rigorous in Ref.~\cite{HC10a}.  
Ref.~\cite{RSBS09} utilized the Bethe-Salpeter equation with a dynamically
screened Coulomb interaction. 
while Ref.~\cite{GB09} extended the original approach of
Ref.~\cite{MZCB04} by taking account of the coupling of the
single-double pair with the entire KS spectrum via the common energy denominator approximation (CEDA).

\subsection{When are double excitations important?}
\label{doub}
Even the low-lying spectra of some molecules are interspersed with
states of double-excitation character, but we will argue that they
also lie at the root of several significant challenges approximate
TDDFT faces for spectra and photo-dynamics. Although not traditionally
seen as a double-excitation problem, we will see that
double-excitations haunt the difficulty in describing conical
intersections and certain long-range charge-transfer states.

{\underline{\it Molecular spectra}}
First, double-excitations in their own right are prominent in the
low-lying spectra of many conjugated polymers. A famous case is the
class of polyenes~(see Ref.\cite{CZMB04} for many references). For example, in
butadiene the HOMO to (LUMO+1) and (HOMO-1) to LUMO excitations are
near-degenerate with a double-excitation of the HOMO to LUMO. If one
runs an adiabatic calculation and simply assigns the energies
according to an expected ordering, one obtains 7.02eV for the vertical
excitation from a B3LYP calculation (similar with other hybrid
functionals), in a 6-311G(d,p) basis set, while a CASPT2 calculation yields 6.27eV. 
By using different basis sets, a more accurate value can appear, but rather fortuitously, since 
the
state obtained in adiabatic B3LYP has more of a Rydberg character, rather
than double character~\cite{HHH01}. In Ref.~\cite{CZMB04}, the dressing
Eq.~(\ref{eq:genspa}), generalized to a subspace of two KS single
excitations instead of one, was applied, yielding 6.28eV. Similar
successes were computed for hexatriene, and also for 0-0
excitations. This system was later studied in detail in
Ref.~\cite{MW09}, analyzing more fully aspects such as self-consistent
treatment of the kernel, and use of KS versus Hartree-Fock
orbitals in the dressing. Further, in Ref.~\cite{MMWA11}, excited state geometries were successfully computed in this way.  Most recently, an extensive study of Eq.~(\ref{eq:genspa})  was performed in Ref.~\cite{HIRC10} to low-lying states of 28 organic molecules.

{\underline{\it Charge-transfer excitations}}
It is well known that long-range charge-transfer excitations are
severely underestimated with the usual approximations of TDDFT. The
usual argument to explain this is that the TDDFT correction to the
bare KS orbital energy difference vanishes because the occupied and
unoccupied orbitals, one being located on the donor and the other on
the acceptor, have negligible overlap as the donor-acceptor distance
increases~\cite{DWH03,T03,A09}. The TDDFT prediction then reduces to the bare KS orbital energy difference, $\epsilon_L({\rm acceptor})- \epsilon_H({\rm donor})$ where the subscripts $H$ and $L$ refer to HOMO and LUMO, respectively. This  typically leads to an underestimation, because in usual approximations $|\epsilon_H|$ underestimates the true ionization energy, while the lowest unoccupied
molecular orbital (LUMO), $\epsilon_{\rm L}$, lacks relaxation
contributions to the electron affinity.  The last few years have seen many  methods to correct the
underestimation of CT excitations, e.g. Refs.~\cite{TTYY04p,SKB09,HIG09}; most
modify the ground-state functional to correct the approximate KS
HOMO's underestimation of $I$, and mix in some
degree of Hartree-Fock, and most, but not all~\cite{SKB09,HIG09} determine this mixing via at least one empirical parameter. 

\begin{figure}[t]
    \includegraphics[width=8.5cm]{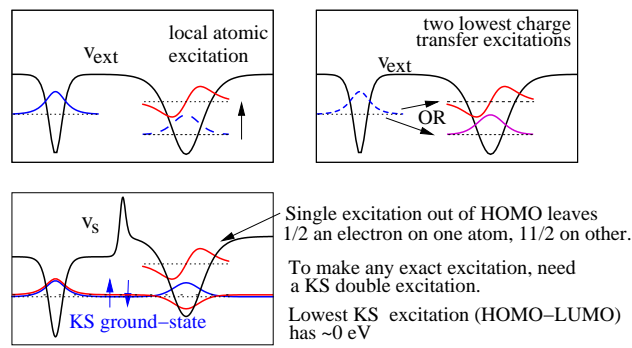}
  \caption{\label{fig:CT} Excitations out of a heteroatomic molecule composed of open-shell fragments at long-range (eg. "stretched" LiH). Blue denotes orbitals occupied in the ground-state; in this model, we show one electron on each "atom" (the inner electrons play only a secondary role).  The top panel shows a model of the possible excitations of the true system: on the left is a local excitation on one atom, on the right are shown two lowest charge-transfer excitations. The bottom left panel shows the corresponding KS potential, displaying a step and a peak, as discussed in the text. The ground-state KS is the doubly-occupied bonding orbital; any single excitation out of here is near-degenerate with a double-excitation where the other electron occupying the bonding orbital transits to the near-degenerate antibonding LUMO.}
\end{figure}

But the argument above only applies to the case
where the donor and acceptor are closed-shell species; instead, if we
are interested in charge-transfer between open-shell species (e.g. in
something like LiH), the HOMO and LUMO are delocalized over the whole
molecule. This is the case for the exact ground-state KS potential, as
well as for semi-local approximations~\cite{TMM09}. The exact KS potential has a peak and a step in the bonding region, that has exactly the size to realign the atomic HOMO's of the two fragments~\cite{P85b,PPLB82,TMM09} (see also 
Fig.~\ref{fig:CT}). As a result the molecular HOMO and LUMO are delocalized over the whole molecule. The HOMO-LUMO
energy difference goes as the tunnel splitting between the two
fragments, vanishing as the molecule is pulled apart; therefore {\it
  every} excitation out of the KS HOMO is near-degenerate with a
double-excitation where a second electron goes from the HOMO to the
LUMO (at almost zero KS cost). This KS double-excitation is crucial to capture the correct nature of the true excitations as 
otherwise we are left with half-electrons on each fragment, physically impossible in the 
dissociated limit. This strongly affects the kernel
structure, imposing a severe frequency-dependence for all excitations,
charge-transfer and local, for heteroatomic molecules composed of
open-shell fragments at large separation~\cite{M05c,MT06}. 

{\underline{\it Coupled electron-nuclear dynamics}}
The importance of double-excitations for coupled electron-nuclear
dynamics was highlighted in Ref.~\cite{LKQM06}:  
even when the vertical excitation does not
contain much double-excitation character, the propensity for
curve-crossing requires accurate double-excitation description for
accurate global potential energy surfaces. The same paper pointed out
the difficulties TDDFT has with obtaining conical intersections: in
one example there, the TDDFT dramatically exaggerated the shape of the
conical intersection, while in another, its dimensionality was wrong,
producing a seam rather than a point. Although a primary task is to correct the ground-state
surface, the problem of
double-excitations is likely very relevant around the conical
intersections due to the near-degeneracy. 

{\underline{\it Autoionizing resonances}}
In the next section we discuss the case when the excitation energy of
a double-excitation is larger than the ionization energy of the
system.  In this case, an autoionizing resonance results.

\section{Autoionizing Resonances in the He atom}
\label{sec:autoionizing}
Photoionization has a rich history in quantum mechanics, with the
photoelectric effect playing a pivotal role in establishing the dual
nature of light, and continues to be a valuable tool in analyzing atoms, molecules, and solids.
The photospectra above the ionization threshold are
characterized by autoionizing resonances, whereby bound state
excitations interact with those into the continuum.
Many theoretical methods\cite{F78,KG88,SM08,SC01,LL00,CMR00} have been developed in order to predict
these resonances.

As TDDFT is both accurate and relatively computationally inexpensive,
it stands as a useful candidate for studying these excitations. For
autoionizing resonances where a single excitation frequency (e.g. core
$\rightarrow$ Rydberg) lies in the continuum, TDDFT has been shown to
work well\cite{SFD05,FSD03,FSD04,STFD06,STFD07,SDL95,HB09}. However it was noted\cite{SDL95,FSD04} that
resonances arising from bound double excitations are completely
missing in the adiabatic approximation (as we might expect from
section \ref{doub}).

Ref. \onlinecite{KM09} derived a frequency-dependent kernel which
allows TDDFT to predict bound-double autoionizing resonances. Below we
review this derivation before testing it on the Helium atom.

\subsection{Frequency-dependent kernel for autoionizing double-excitations}
\label{sec:fano}
Fano's pioneering work on photoionization~\cite{F61} considered a
zeroth-order unperturbed system with a bound state $\Phi_b$ degenerate
with that of a continuum state $\Phi_E$. The unperturbed system
differs from the true system by the coupling term $\hat{V}_{cpl}$, and
we define the matrix element $V_E = \langle\Phi_E |
\hat{V}_{cpl}|\Phi_b\rangle$ between the two states. The transition
probability, for some transition operator $\hat{T}$ (e.g. the dipole),
between an initial state $|i\rangle$ and the mixed state with energy
in the continuum was then found to be
\ben
\label{fano}
\frac{|\langle\Psi_E |\hat{T}|i\rangle |^2}{|\langle\Phi_E |\hat{T}|i\rangle |^2} = \frac{(q+\epsilon)^2}{1+\epsilon^2}
\een
where 
\ben
\epsilon = \frac{E-E_r}{\Gamma/2}
\een
and the energy of the resonance is shifted to 
\ben
E_r = E_b + \mathcal{P}\int dE'\frac{|V_{E'}|^2}{E-E'}
\label{eq:Er}
\een
The asymmetry of the lineshape is given by
\ben
q = \frac{\langle\Phi_b |\hat{T}|i\rangle + \mathcal{P}\int dE' V_{E'}\langle\Phi_{E'} | \hat{T} |i\rangle/(E-E')}{\pi V_{E}\langle\Phi_E |\hat{T}|i\rangle}
\label{eq:q}
\een
while the width of the resonance is given by 
\ben
\label{gam}
\Gamma = 2\pi |\langle\Phi_E | \hat{V}_{cpl}|\Phi_b\rangle|^2 = 2\pi |V_E|^2
\een   

In Ref. \onlinecite{KM09}, this formalism is applied to the KS system, where $\Phi_b$ and $\Phi_E$ are now interpreted as KS wavefunctions, and the full Hamiltonian is related to the KS Hamiltonian via
\ben
\hat{V}_{cpl} = \hat{V}\ee - \hat{v}\H - \hat{v}\xc 
\een
By comparing Eq. (\ref{fano}) with the TDDFT linear response equations, a frequency-dependent kernel, valid in the region near the resonance, was derived
\ben
f\Hxc(\omega) = \chi\s^{-1} - \left( \chi\s + \frac{\Gamma(\Gamma/2 + i(\omega-\omega_r))}{(\omega-\omega_r)^2 +\left(\frac{\Gamma}{2}\right)^2}\Im\chi\s \right)^{-1}
\een
In Eq.~\ref{eq:q}, $q=1$ is required by the facts that both the KS system and the true system respect the Thomas-Reiche-Kuhn sum rule and that the double-excitation does not contribute to the oscillator strength of the KS system~\cite{KM09}. This kernel can be used to 'dress' the absorption spectra found via the adiabatic approximation (AA), $\sigma^A(\omega)$, giving an absorption spectra
\ben
\sigma(\omega) = \frac{(\omega - \omega_r +\Gamma/2)^2}{(\omega-\omega_r)^2+\left(\frac{\Gamma}{2}\right)^2}\sigma^{A}(\omega)
\een
which contains both the AA excitations and the autoionization resonances.

A few points are worth noting. Fano's zeroth order picture is assumed
to take care of all interactions except for the resonance one. When we
approximate this by the non-interacting KS one, we can therefore only
expect the results to be accurate in the limit of weak interaction,
where the dominant interaction is the resonant coupling of the bound
double-excitation with the continuum states.  When the kernel is
included on top of an adiabatic one, that includes also mixing of
non-resonance single excitations, it is done in an a posteriori way,
i.e. non-self-consistently, and expressions for the width are
unaltered, do not include any adiabatic correction. It is also worth
noting that the derivation considers an isolated resonance: just one
discrete state and one continuum.

\subsection{Application: $1s^2\rightarrow 2s^2$ resonance in the He atom}
In order to test the accuracy of this prescription, we studied the $1s^2\rightarrow 2s^2$ autoionization resonance of Helium. This is the lowest double-excitation in the He atom, and it lies in the continuum: Experimentally this excitation occurs at a frequency of $\omega = 57.82~$eV while the ionization threshold is $I = 24.59$~eV.  In this case the wavefunctions are given by
\bea
\Phi_b(\br,\br') &=& \phi_{2S}(\br)\phi_{2S}(\br') \\
\Phi_E(\br,\br') &=& \frac{1}{\sqrt{2}}\left( \phi_{1S}(\br)\phi_{E}(\br') + \phi_{1S}(\br')\phi_{E}(\br)\right) 
\eea
where $\phi_{1S}(\br)$ and $\phi_{2S}(\br)$ are bound KS orbitals while $\phi_{E}(\br)$ is the energy-normalized continuum state with energy $E$. For various xc functionals, the bound state orbitals were calculated using OCTOPUS\cite{octo} while the unbound state was found using an RK$4$ integrator on the KS potential.

In order to produce a continuum state with the correct asymptotic form, the KS potential should decay like $-1/r$. Two functionals which meet this condition for Helium are the local density approximation with the Perdew-Zunger\cite{PZ81} self-interaction correction (LDA-SIC) and the exact exchange functional within the optimized effective potential method (OEP EXX). 

In table \ref{tab:HeGamma}, we show the results for the autoionization
width found using Eq. (\ref{gam}). As can be seen the widths for the
transition to the $2s^2$ state are too low. Closer inspection of the bound-state electronic structure reveals that 
there may be significant mixing between the $2s^2$ state and the $^1${\bf S}
$2p^2$ state.
So we diagonalize the full Hamiltonian in this two-by-two subspace
in order to include
the effect of this configuration interaction mixing. From this
diagonalization, we take the state dominated by $2s^2$ wavefunction
(about $70\%$ in our cases). The autoionization width for this mixed
state (denoted ``two-configuration'') is also shown in Table \ref{tab:HeGamma} where it indeed
improves the pure $2s^2$ state results, but is still roughly a factor
of $40\%$ too small. The LDA-SIC functional performed better than
exact-exchange, this may be due to correlation being strongest in the
core region which contributes most to the integral of Eq. (\ref{gam}). The width for the $2P^2$ excitation is too small to be measured experimentally, however we can compare to other theoretical calculations. After diagonalization, the width does become much smaller compared to the pure state, as it should, however the value is an order of magnitude smaller than that of Ref. \onlinecite{FI90}. 


\begin{table}
\caption{\label{tab:HeGamma} Autoionization widths for Helium as calculated by various functionals.}
\begin{ruledtabular}
\begin{tabular}{cll}
$^1${\bf S} state & Method 					& $\Gamma$(ev) \\
\hline
$2s^2$ 	&									& \\
		& OEP EXX pure 						& $0.0604$ \\
 		& OEP EXX two-configuration 			& $0.0741$ \\ 
		& LDA-SIC pure						& $0.0670$ \\
		& LDA-SIC two-configuration 			& $0.0836$ \\
		& MCHF+b-spline\footnotemark[1]		& 0.1529 \\
		& Experiment\footnotemark[2]				& $0.138\pm 0.015$ \\
\hline
$2p^2$ 	&									& \\ 					
		& OEP EXX pure						& $0.0139$ \\	
 		& OEP EXX two-configuration 			& $0.0003$ \\  			
  		& LDA-SIC pure						& $0.0170$ \\		
  		& LDA-SIC two-configuration 			& $0.0004$ \\  			
		& MCHF+b-spline\footnotemark[1]		& 0.0055 \\
		& Experiment						& --\\				
\end{tabular}
\end{ruledtabular}
\footnotemark[1]{From Ref.\cite{FI90} }
\footnotemark[2]{From Ref.\cite{HC75} }
\end{table}

In conclusion, we tested the formalism of Ref.~\cite{KM09} to
include double-excitation autoionization resonances within TDDFT for
the Helium atom. The results, while not outlandish, are disappointing
when compared to simpler wavefunction methods\cite{MF76} for this resonance in
Helium. This is probably due to the fact that the KS system in this case does not make a good zeroth-order picture on which to build a Fano formalism: i.e. the
assumption of weak interaction, aside from the resonance coupling, discussed in Sec.~\ref{sec:fano} do not hold. 
It would be interesting to compute the shift in the resonance position (Eq.~(\ref{eq:Er})) within the TDDFT prescription; this is expected to be large, e.g. comparing 
the KS double ($\omega =
40.6$~eV in exact TDDFT\cite{UG94}) to the experimental resonance ($\omega =
57.82~$eV). Two approximate functionals were tested, LDA-SIC and exact
exchange, with the results suggesting that correlation improves the
description. Functionals missing the $-1/r$ tail in their KS
potentials can provide continuum states accurate within a core
region\cite{WMB05,WB05} (although missing the correct asymptotic behavior) and
so may still be used with this method. However they are unlikely to significantly improve the results for the reasons discussed above.

Hellgren and von Barth derived the Fano lineshape formula in terms of
linear response quantities, that is exact within the adiabatic
approximation~\cite{HB09}. It cannot apply to the case of a bound
double-excitation. On the other hand, our expression does apply but it is a lower-order approximation, only expected to be accurate in the limit of a narrow isolated resonance in the weak interaction limit. 
For molecules and
larger systems, TDDFT is often the only available technique to
calculate excitations and this formalism is the only available method
to treat doubly-excited autoionization resonances within TDDFT. For
such larger systems, we expect this approach might provide useful
results.

\section{Adiabatic quadratic response: double vision?}
\label{sec:quadratic}
Given that to excite two electrons in a non-interacting systems, two
photons are required, one may ask whether adiabatic TDDFT in {\it
  nonlinear} response, especially quadratic response, yields
double-excitations.  In this section, we work directly with the
second-order KS and interacting response functions to investigate this
question. We will first consider the form of the non-interacting quadratic response function and briefly review TDDFT response theory, before turning specifically to the question of the double-excitations. 

Applying a perturbation to a system $\delta v(\br,t)$ initially in its
ground state $n_0(\br)$, we may expand the response of the density in
orders of $\delta v(\br,t)$, where $n(\br,t) = n_0(\br) + n_1(\br,t) +
n_2(\br,t) + ...$. 
We define corresponding response functions:
\bea
\chi(\br,t, \br',t') &=& \left.\frac{\delta n[v](\br,t)}{\delta v(\br',t')}\right\vert_{v = v_0} \nonumber \\
&=& \theta(t - t')\langle \Psi_0\vert [\hat{n}_H(\br,t),\hat{n}_H(\br',t')]\vert \Psi_0\rangle \nonumber \\
& &\label{eq:chiagain}
\eea
as the linear-response density-density response function whose frequency-domain version appeared earlier in Eqs.~(\ref{eq:Dysonlin}-\ref{eq:chi}), 
and the second-order response function
\bea
\nonumber
\chi^{(2)}(\br,t, \br',t', \br'',t'') = \left.\frac{\delta^2 n[v](\br,t)}{\delta v(\br',t')\delta v(\br'',t'')}\right\vert_{v = v_0} & &\\
\nonumber
= -\half\left\{ \langle \Psi_0\vert [[\hat{n}_H(\br,t),\hat{n}_H(\br',t')],\hat{n}_H(\br'',t'')]\vert \Psi_0\rangle \right. & & \\
\left. + {\rm permutations\; of} (\br',t') \leftrightarrow (\br'',t'')\right\} \theta(t - t')\theta(t - t'')& & \nonumber \\
\label{eq:chi2}
\eea
where $\hat{n}_H$ is the density operator in the Heisenberg picture and the expressions in terms of the commutators follow from time-dependent perturbation theory~\cite{GDP96, WH74}.

In the following we often will drop explicit spatial-dependences, and
think of density as a vector. Kernels and response functions would then be represented as matrices and tensors with the symbol $\star$ signaling contraction.

\subsection{The non-interacting quadratic response function}
\label{sec:nonintquad}
First, we ask, what is happening at the level of the non-interacting
KS response functions? In the above equations, the functional derivatives with respect to $v$ are replaced by those with respect to the corresponding $v\s$ and the ground-state in Eqs.~(\ref{eq:chiagain}-\ref{eq:chi2}) is then a single Slater determinant.
Expanding out the commutator in Eq.~(\ref{eq:chi2}) and inserting the identity in the form of
completeness relations of the Slater-determinant basis, we obtain
terms of the form
\ben
\sum_{I\s}\sum_{J\s} \langle 0\s\vert \hat{n}(\br) \vert I\s \rangle \langle I\s\vert \hat{n}(\br') \vert J\s \rangle  \langle J\s\vert \hat{n}(\br'') \vert 0\s \rangle 
\een
where the subscript $S$ indicates it's a KS state. Examining the first
bracket, we see that $I\s$ must be either the ground-state or a
single-excitation, since the one-body density operator can only
connect determinants differing by at most one orbital. Likewise, for a
non-zero third bracket, $J\s$ can only be the ground-state or a
single-excitation. Therefore, no double-excitations contribute to the
second-order response function.  At second-order, double-excitations
are reached in the dynamics of a non-interacting system but cannot
contribute to the one-body {\it density} response.
By similar arguments, the lowest order density-response that double-excitations can appear is at third-order.
That the KS second-order response function does not contain
double-excitations begins to dash any hope that adiabatic TDDFT will yield
accurate excitations. It will turn out that we do ``see double'', but the
vision is blurry: they are simply sums of linear-response corrected
single-excitations, quite blind to any nearby single
excitation. Before turning to a closer investigation of this, we
briefly review the structure of TDDFT response theory.

\subsection{Non-linear response in TDDFT}
The fact that the time-dependent KS system reproduces the true system's
density response to the corresponding KS perturbing potential $\delta
v\s(\br,t)$, leads to Eq.~(\ref{eq:Dysonlin}) for
the linear response function, and~\cite{GDP96}
\begin{widetext}
\bea
\chi^{(2)}(\br,t, \br',t', \br'',t'') &=& \int ds~d^3y\int ds'~d^3y' \chi\s^{(2)}(\br,t, \by,s, \by',s')
\left.\frac{\delta v\s(\by,s)}{\delta v(\br',t')}\right\vert_{n_0}\left.\frac{\delta v\s(\by',s')}{\delta v(\br'',t'')}\right\vert_{n_0} \nonumber \\
&+& \int ds~d^3y \chi\s(\br,t, \by,s) \int ds'~d^3y'\int ds''~d^3y'' g\xc[n_0](\by,s,\by',s',\by'',s'') \chi(\by',s',\br',t') \chi(\by'',s'', \br'',t'') \nonumber \\
&+& \int ds~d^3y \chi\s(\br,t, \by,s) \int ds'~d^3y' f\Hxc[n_0](\by,s,\by',s') \chi^{(2)}(\by',s',\br',t',\br'',t'')
\eea
\end{widetext}
for the quadratic response function. Here $\chi\s^{(2)}$ denotes the second-order KS response function, discussed in Sec.~\ref{sec:nonintquad} and
\ben
g\xc[n_0](\br,t,\br',t',\br'',t'')=\left.\frac{\delta^2 v\xc[n](\br,t)}{\delta n(\br',t')\delta n(\br'',t'')}\right\vert_{n=n_0}
\een
 is the dynamical second-order xc kernel. For simplicity we assume the state is spin-saturated. In the adiabatic approximation, $g\xc^A(\br,\br',\br'') = \frac{\delta^3E\xc[n]}{\delta n(\br)\delta n(\br') \delta n(\br'')}$ with $E\xc[n]$  a ground-state energy functional. 
Making a Fourier transform with respect to $t-t'$ and $t-t''$ we obtain the first and second-order density responses as
\bea
\nonumber
\n_1(\omega) &=& \chi(\omega) \star \delta v(\omega) \\
&=& \chi\s(\omega)\star \delta v\s(\omega) \nonumber \\
&=& \chi\s(\omega)\star \delta v(\omega) + \chi\s(\omega)\star f\Hxc(\omega)\star\n_1(\omega) \nonumber \\
\label{eq:n1}
\eea
and 
\bea
\nonumber
n_2(\omega) &=& \frac{1}{2}\int d\omega' \chi^{(2)}(\omega,\omega-\omega') \star \delta v(\omega)\delta v(\omega-\omega')  \\
&=& \half\int d\omega' \chi\s^{(2)}(\omega,\omega-\omega')\star \delta v\s(\omega) \delta v\s(\omega-\omega')  \nonumber \\
+ \half \ \chi\s(\omega)&\star & g\Hxc(\omega,\omega-\omega')\star\int d\omega'  \n_{1}(\omega') \n_{1}(\omega-\omega')  \nonumber \\
+ \chi\s(\omega) &\star & f\Hxc(\omega) \star\n_2(\omega) 
\label{eq:n2}
\eea
respectively.

Ref.~\cite{GDP96} pointed out a very interesting structure that the TDDFT response equations have. At any order $i$, 
\ben
n_i(\omega) = M_i(\omega)+ \chi\s(\omega)\star f\Hxc(\omega) \star n_i(\omega)
\label{eq:ithresp}
\een
where $M_i$ depends on {\it lower-order} density-response (and response-functions up to $i$th order). The last term on the right of Eq.~\ref{eq:ithresp} has the same structure for all orders. If we define the operator 
\ben
L(\omega) = \mathds{1} - \chi\s(\omega) \star f\Hxc
\label{eq:L}
\een
then
\ben
L(\omega)\star n_i(\omega) = M_i(\omega)
\label{eq:ithresp2}
\een

In the next section we work out the effect of this operator in the
adiabatic approximation, by studying linear response. This will be
useful to us when we finally ask about doubles in the quadratic
response function.

\subsection{Adiabatic approximation}
\label{sec:adiabL}
Let us find the inverse of the operator appearing on the left-hand-side
of Eq.~\ref{eq:ithresp2}. First, evaluating Eq.~\ref{eq:chi} with KS determinants and KS energies, one finds that, for $\omega$ not equal to a KS transition frequency or its negative, 
\ben
\chi\s(\omega) = \sum_q \frac{A_{{\sss S},q}}{\omega^2-\omega_q^2}
\een
where $q$ labels a transition from occupied orbital $i$ to an unoccupied orbital $a$ and the matrix
\ben
A_{{\sss S},q}(\br,\br') = 4\omega_q \phi_i(\br)\phi_a(\br)\phi_a(\br')\phi_i(\br')
\een
where $\omega_q = \epsilon_a - \epsilon_i$ and we take the KS orbitals to be real. We note that there is a factor of 2 in $A_{{\sss S},q}(\br,\br')$ due to the assumed spin-saturation. 
Also note that if only forward transitions are kept, this amounts to what is often called the Tamm-Dancoff approximation:
\ben
\chi\s(\omega) = \sum_q \frac{A_{{\sss S},q}^{TD}}{\omega-\omega_q}
\een
with $A_{{\sss S},q}^{TD} = 2\phi_i(\br)\phi_a(\br)\phi_a(\br')\phi_i(\br')$.

Returning to Eq.~\ref{eq:ithresp2}, for an adiabatic approximation, $f\Hxc(\omega) = f\Hxc^A$, we may write
\ben
\chi\s(\omega) \star f\Hxc^A = \sum_q \frac{\left(A_{\sss S}\star f\Hxc^A\right)_q}{\omega^2 - \omega_q^2}
\label{eq:chifhxc}
\een
where $A_{\sss S}\star f\Hxc^A = 4\omega_q \phi_i(\br)\phi_a(\br)\int f\Hxc^A(\br_1,\br')\phi_i(\br_1)\phi_a(\br_1)d^3r_1$.

This means the effect of the operator on the left-hand-side of
Eq.~\ref{eq:ithresp2}, is to zero out poles of the function it is
operating on that lie at KS single excitations, and replace them with
linear-response-corrected ones, within the adiabatic approximation. In particular, 
\begin{widetext}
\ben
L^{-1}(\omega) =\prod_q(\omega^2 - \omega_q^2)\left(\prod_q(\omega^2 - \omega_q^2)\mathds{1} - \sum_q(A_{\sss S}\star f\Hxc^A)_q \prod_{q' \neq q}(\omega^2 - \omega_{q'}^2)\right)^{-1}
\label{eq:Linv}
\een
\end{widetext}
(where $\Pi_q$ indicates a product over $q$)
It is instructive to first consider linear response ($i=1$ in Eq.~\ref{eq:ithresp2}), 
and zoom in on only one excitation, with coupling to all others considered insignificant. 
Then, from Eqs.~\ref{eq:chifhxc},~\ref{eq:L},~\ref{eq:Linv}, and~\ref{eq:n1}
\bea
\nonumber
n_1(\omega)&=& L^{-1}(\omega)\star\chi\s(\omega)\star \delta v(\omega)\\
&\approx&  \left((\omega^2 -\omega_q^2)\mathds{1} - (A_{\sss S}\star f\Hxc^A)_q\right)^{-1}\star A_{{\sss S},q}\star \delta v(\omega) \nonumber \\
\label{eq:Linvchis}
\eea
Poles of $n_1(\omega)$ are thus indicated by zeroes of the first term in brackets: when $(\omega^2 -\omega_q^2)\delta(\br - \br') - (A_{\sss S}\star f\Hxc^A)_q = 0$. Integrating this over $\br_2$ and realizing the second term gives zero unless $\br' = \br$, we find we have effectively derived the ``small-matrix approximation''~\cite{VOC99, AGB03}
\ben
\omega^2 = \omega_q^2 + 4\omega_q\int\phi_i(\br)\phi_a(\br)f\Hxc^A(\br,\br')\phi_i(\br')\phi_a(\br') d^3rd^3r'
\een
Had we kept only forward terms, and proceeded in a similar manner, the
single-pole approximation (Eq.~\ref{eq:spa} but with a
frequency-independent right-hand-side) would have resulted.
The essential point is that the effect of $L^{-1}$ is to shift the KS pole towards the linear-response corrected single excitation. 

Now this operator appears at all orders of TDDFT response, yielding
critical corrections to the non-interacting response functions at all
orders. We now are ready to investigate the question of double-excitations in quadratic response in the adiabatic approx.


\subsection{The quadratic density-response in the adiabatic approximation}
We have from Eq.~(\ref{eq:n2}) and definition~(\ref{eq:L}) that
\begin{widetext}
\label{eq:Ln2}
\ben
\n_2(\omega) = \half\left(L^{-1}(\omega)\star\int d\omega' \chi\s^{(2)}(\omega-\omega',\omega)\star \delta v\s(\omega-\omega')\star \delta v\s(\omega') 
+  L^{-1}(\omega)\star \ \chi\s(\omega)\star g^A\Hxc\star\int d\omega'  \n_{1}(\omega') \n_{1}(\omega-\omega')\right)
\een
\end{widetext}
 We wish to investigate specifically, the question of how double-excitations would appear in the poles of $n_2(\omega)$. 
From Sec.~\ref{sec:adiabL}, we know the effect of the operator $L^{-1}(\omega)$ acting on a function with a pole at a KS single excitation, is to shift that pole to its linear-response corrected value, in the adiabatic approximation, (Eqs.~(\ref{eq:Linv} and (\ref{eq:Linvchis}). Let us denote these values as $\Omega_I$. 
We shall now study in detail the  second term in Eq.~\ref{eq:Ln2}, which is
\begin{widetext}
\ben
\prod_q(\omega^2 - \omega_q^2)\left(\prod_q(\omega^2 - \omega_q^2)\mathds{1} - \sum_q(A_{\sss S}\star f\Hxc^A)_q \prod_{q'' \neq q}(\omega^2 - \omega_{q''}^2)\right)^{-1} \star \sum_{q'}\frac{(A_{\sss S}\star g\Hxc^A)_{q'}}{\omega^2-\omega_{q'}^2} \star\int d\omega'  \n_{1}(\omega') \n_{1}(\omega-\omega')
\een
\label{eq:secondterm}
\end{widetext}

So until the last integral over $\omega'$, only poles at single KS excitations corrected by linear-response adiabatic TDDFT appear. We now turn to this last integral to see what it gives. First, we write it as 
\ben
\int d^3y d^3y' \int d\omega'  \chi(\br,\by,\omega')\chi(\br',\by',\omega-\omega') G(\by,\by',\omega',\omega-\omega')
\label{eq:lastintegral}
\een
where $G(\by,\by',\omega',\omega-\omega') = v_1(\by,\omega')v_1(\by',\omega-\omega')$. After applying an adiabatic kernel in Eq. (\ref{eq:Dysonlin}), the linear response function can be written as
\ben
\chi^{A}(w) = \sum_I \left( \frac{A_I}{\omega-\Omega_I+i0_+} - \frac{A_I}{\omega+\Omega_I+i0_+} \right)
\een
which contains the same number of poles as $\chi\s(\omega)$, in contrast to Eq. (\ref{eq:chi}). Inputting this adiabatic linear-response function into Eq.~(\ref{eq:lastintegral}), we find:
\begin{widetext}
\ben
\int d\omega' \sum_I \sum_J A_I A_J \left( \frac{1}{\omega'-\Omega_I} - \frac{1}{\omega'+\Omega_I} \right) \left( \frac{1}{\omega'-\omega-\Omega_J} - \frac{1}{\omega'-\omega+\Omega_J} \right)G(\omega',\omega-\omega')
\een

Doing the integrals, we obtain
\bea
= 2\pi i  \sum_I \sum_J A_I &A_J &\left( \frac{G(\omega+\Omega_J,-\Omega_J)-G(\Omega_I,\omega-\Omega_I)}{\omega-(\Omega_I-\Omega_J)} + \frac{G(\Omega_I,\omega-\Omega_I)G(\omega-\Omega_J,\Omega_J)}{\omega-(\Omega_I+\Omega_J)} \right. \nonumber \\
&+& \left. \frac{G(-\Omega_I,\omega+\Omega_I)-G(\omega+\Omega_J,-\Omega_J)}{\omega+(\Omega_I+\Omega_J)} + \frac{G(\omega-\Omega_J,\Omega_J)-G(-\Omega_I,\omega+\Omega_I)}{\omega+(\Omega_I-\Omega_J)} \right)
\eea
\end{widetext}
This displays poles at sums and differences of the linear-response-corrected single-excitations.  In particular, it contains poles at double-excitations, $\pm (\Omega_I +\Omega_J)$.  These poles remain after being multiplied by the preceding terms in Eq.~\ref{eq:secondterm} (which, as explained before, may cancel/shift a pole at a single KS excitation).
Notice that if we made a Tamm-Dancoff-like approximation for $\chi$ with respect to $\omega'$ then Eq.~\ref{eq:lastintegral} would be
\ben
 -\sum_{IJ} A_I A_J \int d\omega' \left( \frac{1}{\omega'-\Omega_I} \right) \left( \frac{1}{\omega'-\omega-\Omega_J} \right) G(\omega',\omega-\omega')\;,
\een
yielding
\ben
 \sum_{IJ} A_I A_J \frac{G(\Omega_I,\omega-\Omega_I)-G(\omega+\Omega_J,-\Omega_J)}{\omega-(\Omega_I-\Omega_J)}
\een
That is, the double excitations vanish in this Tamm-Dancoff-like approximation.

Our findings here are not entirely new: in Ref.~\cite{TC03} very
similar conclusions were reached, but by quite a different method. To our
knowledge, an analysis based directly on the response functions, as
above, has not appeared in the literature before. The formalism in
Ref.~\cite{TC03} was based on the dynamics of a system of weakly
coupled classical harmonic oscillators, in a coordinate system defined
by the transition densities. The linear and non-linear response of
this system was shown to correspond to that of a real electronic
system via adiabatic TDDFT. Closed expressions for the first, second,
and third optical polarizabilities were found; in particular the
second-order polarizability was shown to contain poles at sums of
linear-response corrected single-excitations, and that if a
Tamm-Dancoff approximation was made, these poles vanish. We have shown
very similar results staying within the usual language of TDDFT and
response theory, directly from analyzing the response functions. The
relation between the Tamm-Dancoff-like approximation we make here and
that referred to in~\cite{TC03} remains to be examined in more detail.

So we have shown that the quadratic response in TDDFT within the adiabatic approximation is unable to describe double excitations. While the second-order response in adiabatic TDDFT does contains poles at the sum of linear-response corrected KS frequencies, it completely misses the mixing between single and double KS states needed for an accurate description of double excitations. Moreover, using the Tamm-Dancoff approximation in linear response makes even these poles disappear. These conclusions were reached by looking at the pole structure of the response equations directly. Using this approach, we were able to see that the second-order Kohn-Sham response function does not contain poles at the sum of the KS frequencies, which preventing TDDFT from introducing the mixing mentioned above.

\section{Double-excitations via semiclassical dynamics of the density-matrix}
\label{sec:frozengauss}
Recently, time-dependent density-matrix functional theory (TDDMFT) has
been explored as a possible remedy for many of the challenges of
TDDFT~\cite{PGB07p,GBG08p}. The idea is that including more information in
the basic variable would likely somewhat
relieve the job of xc functionals and lead to simpler functional
approximations working better. The (spin-summed) one-body density-matrix is
\begin{widetext}  
\ben
\rho(\br',\br,t)= N\sum_{\sigma_1}\int dx_2..dx_N
\Psi^*(\br' \sigma_1,x_2..x_N,t)\Psi(\br \sigma_1,x_2..x_N,t)
\een
\end{widetext}
(using $x= (\br,\sigma)$ as spatial-spin index with $\int dx = \sum_\sigma\int d^3r$), 
 while the density is only the diagonal
element, $n(\br,t) = \rho(\br,\br,t)$.
For example, immediately we see that
the kinetic energy is exactly given by the one-body density-matrix as $T = -\int d^3r \frac{1}{2}\nabla^2 \rho(\br'\br,t)\vert_{\br'=\br}$,
while only approximately known as a functional of the density alone; and,
only the non-interacting part of the kinetic energy is directly
calculated from the KS orbitals (as $-\sum_i\int dx \phi_i^*(x)\nabla^2\phi_i(x)/2$), while it is unknown how to extract the
exact interacting kinetic energy from the KS system. TDDMFT works
directly with the density-matrix of the interacting system so is not
restricted to the single-Slater-determinant feature of the TDKS system.
For this reason, (TD)DMFT would be especially attractive for
strongly-correlated systems, for example dissociating molecules and it
was recently shown that adiabatic approximations in TDDMFT are able to
capture bond-breaking and charge-transfer excitations in such
systems~\cite{GBG08p}. However, it was also shown that adiabatic
approximations within TDDMFT still cannot capture
double-excitations~\cite{GBG08p}.

Very recently a semiclassical approach to correlation in TDDMFT has
been proposed~\cite{RRM10}, which was argued to overcome several failures that
adiabatic approximations in either TDDFT or TDDMFT
have. The applications in mind involved real-time
dynamics in non-perturbative fields, for example, electronic quantum
control via attosecond lasers, or ionization processes. In these
applications, memory-dependence, including initial-state dependence,
is typically important~\cite{RGHM09,M05b}, but lacking in any
adiabatic approximation. More severely, the issue of time-evolving
occupation numbers becomes starkingly relevant: typically, even when a
system begins in a state which is well-approximated by a single-Slater
determinant, it will evolve to one which fundamentally involves more
than one Slater determinant (eg. in quantum control of He from $1s^2
\to 1s2p$, or in ionization)~\cite{RGHM09}. Although impossible when
evolving with one-body Hamiltonians such as in TDKS (thus making the
job of the exact xc potential very difficult, as well as observables
to extract information from the KS system), in principle TDDMFT can
change occupation numbers. However it was recently proven that
adiabatic approximations in TDDMFT cannot~\cite{RP10,AG10,GGB10}.  The
semiclassical correlation approach of Ref.~\cite{RRM10} incorporates
memory, including initial-state dependence, and does lead to
time-evolving occupation numbers, as has been demonstrated on model
systems~\cite{EM11}.  We now ask, does it capture double-excitations
accurately? We shall use a model two-electron system to investigate
this question, but first will review the method. Unlike the previous sections in the paper, this operates in the real-time domain, instead of the frequency-domain. 

\subsection{Semiclassical correlation in density-matrix propagation}
\label{sec:semiclassicalcorrelation}
The equation of motion of $\rho$ is given by 
\bea
i\dot{\rho}(\br',\br;t) &=& \left( -\frac{\nabla^2}{2} + v\ext(\br;t) +\frac{\nabla'^2}{2} - v\ext(\br') \right)\rho(\br',\br;t) \nonumber \\\
& &+ \int d\br_2 f\ee(\br',\br,\br_2)\rho_2(\br',\br_2,\br,\br_2;t) \label{eq:rho1dot}
\eea
where $f\ee(\br'\br,\br_2) = 1/|\br-\br_2| - 1/|\br'-\br_2|$ and $\rho_2$ is the
second-order reduced density matrix defined by:
\ben
\rho_2(\br_1',\br_2',\br_1,\br_2;t) = \sum_{\sigma_1,\sigma_2} \rho_2(\br_1'\sigma_1,\br_2'\sigma_2,\br_1\sigma_1,\br_2\sigma_2;t)
\een 
and
\begin{widetext} 
\ben
\label{eq:rho2}
\rho_2(x_1',x_2',x_1,x_2;t) = \frac{N(N-1)}{2}\int dx_3..dx_N
\Psi^*(x_1',x_2',x_3..x_N;t)\Psi(x_1,x_2,\br_2\sigma_2,x_3..x_N;t)
\een 
\end{widetext}
It is convenient to decompose this as
\bea
\label{rhoc}
\rho_2(x_1',x_2',x_1,x_2;t) &=&\rho(x_2',x_2;t)\rho(x_1',x_1;t)  \nonumber \\
&-& ~ \rho(x_1',x_2;t)\rho(x_2',x_1;t)  \nonumber \\
&+& ~  \rho\2c(x_1',x_2',x_1,x_2;t) 
\eea
where we identify the first term as the Hartree piece, the second as exchange, and the last as correlation. If $\rho\2c$  is set to zero, one obtains Hartree-Fock; it is this term that Ref.~\cite{RRM10} proposed to treat semiclassically in order to capture memory-dependence and time-evolving occupation numbers~\cite{RRM10}. We shall shortly see that, in contrast to the adiabatic approximations of this term, its semiclassical treatment also approximately captures double-excitations.

There are various different semiclassical formulations, and
the one we will explore here is known as ``Frozen Gaussian'' propagation, proposed
originally by Heller~\cite{H81}. This can be expressed mathematically as a simplified version of the 
Heller-Herman-Kluk-Kay (HHKK) propagator\cite{KHD86,K05} where 
the wavefunction at time $t$: 
\ben
\label{frozg}
\Psi_t(\mathbf{x}) = \int\frac{d\bq_0d\bp_0}{(2\pi\hbar)^N}\langle\mathbf{x}|\bq_t\bp_t\rangle C_{\bq,\bp,t}e^{iS_t/\hbar}\langle\bq_0\bp_0 |\Psi_i\rangle
\een
where $\{\bq_t,\bp_t\}$ are classical phase-space trajectories in $6N$-dimensional phase-space, starting from initial points $\{\bq_0,\bp_0\}$, and $\Psi_i $ is the initial state. In Eq.~\ref{frozg}, $\langle\mathbf{x}|\bq\bp\rangle$ denotes the coherent state: 
\ben
\langle\mathbf{x}|\bq\bp\rangle = \prod_{j=1}^{6N}\left(\frac{\gamma_j}{\pi}\right)^{1/4}e^{-\frac{\gamma_j}{2}(x_j-q_j)^2 + ip_j(x_j-q_j)/\hbar}
\een
where $\gamma_j$ is a chosen width parameter. 
$S_t$ is the classical action and $C_{\bq,\bp,t}$ is a pre-factor
based on the monodromy (stability) matrix. The pre-factor is time-consuming to
compute, and scales cubically with the number of degrees of freedom, but in the Frozen Gaussian approximation, is set to unity.
Ref.~\cite{RRM10} gave an expression for the second-order density-matrix within a Frozen-Gaussian approximation, that, furthermore, takes advantage of the fact that there is some phase-cancellation between $\Psi$ and $\Psi^*$ in Eq.~\ref{eq:rho2} so that the resulting phase-space integral is less oscillatory.

The idea is to extract the correlation term from the semiclassical
dynamics and insert it as a driving term in
Eq.~(\ref{eq:rho1dot})~\cite{RRM10}.  That is, we compute the
semiclassical first-order and second-order reduced-density matrices from Frozen Gaussian dynamics, placing them in Eq.~(\ref{rhoc}) which is inverted to solve for $\rho\2c^{\rm SC}$. As the other terms of
Eq.~(\ref{rhoc}) and~(\ref{eq:rho1dot}) are given exactly in terms of the one-body
density-matrix, we use only the semiclassical expression for $\rho\2c$
when driving Eq.~(\ref{eq:rho1dot}).

It is interesting however to first ask how well semiclassical dynamics alone
does. That is, {\it without} coupling to the exactly-computed
one-body, Hartree and exchange-terms in Eq.~(\ref{eq:rho1dot}), how
would the semiclassical calculation alone predict the dynamics? In
particular, do we obtain double-excitations, and if so, how well.

\subsection{Double excitations from semiclassical dynamics}
We consider the following one-dimensional model of a two-electron
quantum dot:
\ben
\hat{H}= \sum_{i=1,2}\left(-\frac{1}{2}\frac{d^2}{dx_i^2}+\half x_i^2 \right) +
\frac{1}{\sqrt{(x_1-x_2)^2 +1}} 
\een

using a soft-Coulomb interaction between the electrons.
Starting from an arbitrary initial state, the interacting dynamics in
this Hamiltonian can be numerically solved exactly, and we will
compare exact results with those computed by Frozen Gaussian dynamics.

\begin{figure}[t]
    \includegraphics[width=8.5cm]{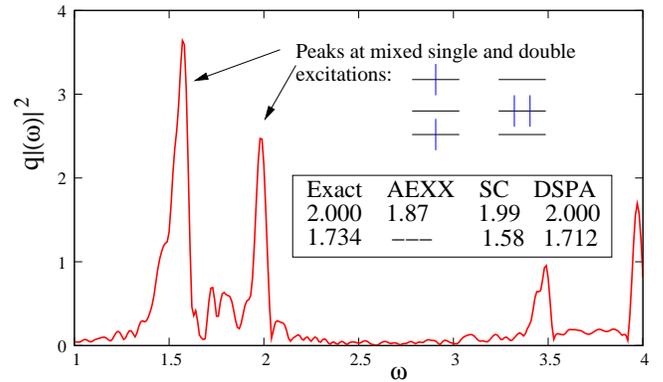}
  \caption{\label{FG_qspec} The Fourier transform of the quadrupole
    moment computed via semiclassical (Frozen Gaussian) dynamics in
    the Hooke's quantum dot. We focus on the region of the spectrum
    where the excitations shown in the level sketch to the upper right
    lie as explained in the text. The table shows the values of the
    excitations computed exactly, using adiabatic exact-exchange
    (AEXX), semiclassical (SC), and the dressed correction of
    Eq.~(\ref{eq:genspa})-(\ref{eq:spa}) (DSPA).  }
\end{figure}

First, by considering a non-interacting reference, we can identify
where single and double excitations lie, which, when interaction is
turned on, will mix. The level sketch in the upper right of
Fig.~\ref{FG_qspec} shows this: in the ground-state, both
electrons occupy the lowest level, and shown is a single excitation to
the second-lowest excited orbital (left) and a double-excitation to
the first excited orbital (right). The non-interacting energies of
these two states are near-degenerate, and mix strongly to give roughly
50:50 single:double mixtures for the true interacting states.
Due to quadratic symmetry of the Hamiltonian, these states do not
appear in the dipole response of the system, hence we look at the quadrupole moment:
\ben
q(t) = \int dx ~  x^2 n(x;t)
\een
We start in an initial state quadratically 'kicked' from the ground state, that is
\ben
\Psi_i(x_1,x_2) = e^{i\eta(x_1^2+x_2^2)}\Psi_0(x_1,x_2)
\een 
where $\eta $ is chosen large enough to sufficiently populate the
states we are interested in without being too large leading to
higher-order response effects. A value of $\eta=0.01$ was used in our
calculation. The Frozen Gaussian integral of Eq. (\ref{frozg}) is
performed by Monte-Carlo integration using trajectories based on
importance sampling the initial state overlap with the coherent
state, and a width of $\gamma = 1$ was used. In the example discussed below, $120000$ trajectories were used, with the quadrupole moment not changing significantly if more trajectories are used.
These trajectories are then classically propagated forward in time using a
standard leapfrog algorithm. A total time of $T=200$au was performed
with a timestep of $\Delta t = 0.001$, although the wavefunction is
only constructed every $0.1$au as this is sufficient to see the
frequencies we are interested in.

In Fig.~\ref{FG_qspec}, we show the power spectrum of the Fourier
transform of the quadrupole moment for a Frozen Gaussian calculation
with the parameters given above. Also given in Fig. \ref{FG_qspec} is
a table comparing the exact, adiabatic exact-exchange (AEXX),
semiclassical (SC), and the dressed correction (DSPA) discussed in the
introduction. As stated earlier, an adiabatic approximation cannot
increase the number of poles in the KS response function: AEXX can
only shift the KS single excitation, yielding a solitary peak in
between the two exact frequencies\cite{TK09}. In contrast to this, the Frozen
Gaussian semiclassical results can be seen to give two peaks in the
right region. Although one peak is lower than in the exact case, this
error may be lessened when $\rho\2c$ is used to drive the
density-matrix propagation, with the one-body terms treated exactly quantum-mechanically instead of semiclassically (Sec.~\ref{sec:semiclassicalcorrelation}). The $\omega=2$ peak arises from excitation in the center-of-mass coordinates where the reduced Hamiltonian is harmonic, given that the Frozen gaussian method is exact for such systems, we would expect this peak to be very accurate.
The DSPA works extremely well in this
case, although in more complicated systems one must search the KS
excitation spectrum in search of nearby doubles, as explained in the
introduction, whereas they will appear naturally in the spectrum in
the semiclassical approach.

This example demonstrates that semiclassical correlation does capture
double-excitations approximately, unlike any adiabatic approach in
TDDFT or TDDMFT.  We stress that this is the result from semiclassical
dynamics alone (within the frozen gaussian approximation); in future
work we will investigate whether the coupling to the exact one-body
terms of Eq.~\ref{eq:rho1dot} improve the results for the excitations.

\section{Summary and Outlook}
TDDFT is in principle an exact theory based on a single-particle reference: the exact functionals extract from the non-interacting KS system, the exact excitations and dynamics of an interacting electronic system. As such, it is both fundamentally and practically extremely interesting how these functionals must look when describing states of double-excitation character, particularly in linear response where no double-excitations occur in a non-interacting reference. The work of Ref.~\cite{MZCB04} that shows the form of the exact xc kernel and models an approximate practical frequency-dependent kernel based on this, has recently drawn some interest, both from a theoretical and practical point of view. We have discussed here how double-excitations are at the root of some of the most difficult problems in TDDFT today: long-range charge-transfer excitations between open-shell fragments, and conical intersections. 

The paper then described three new results in three different approaches involving double-excitations.  First, we have applied a recently proposed approach~\cite{KM09} to autoionizing resonances arising from double-excitations, to compute the width of the 2s$^2$ resonance in the He atom. Although the results are not very accurate, predicting a $40\%$ too narrow resonance, this approach is the only available one for this kind of resonance in TDDFT today.  For larger systems, where the alternative wavefunction-based methods are not feasible, the approach of Ref.~\cite{KM09} might still be useful, despite  the weak-interaction assumption (that is likely responsible for the error in the He case). 
Second, we showed that use of the adiabatic approximation in quadratic response theory yields double-excitations which are the sums of linear-response-corrected single-excitations. Within a Tamm-Dancoff-like approximation, even these doubles disappear. We argued that the KS quadratic response function does not have poles at KS double-excitations, which is behind the reason that we do not see the truly mixed single and double excitations in the adiabatic TDDFT quadratic response function. Although similar conclusions were reached in Ref.~\cite{TC03}, our analysis here proceeds in a very different manner: here, we follow more traditional response theory within DFT, without introducing new formalism. 
Finally, we investigated whether and how accurately double-excitations appear in a recently proposed semiclassical approach to correlation. This approach was originally proposed for general real-time dynamics~\cite{RRM10}, based on propagation of the one-body-density-matrix. The correlation-component of the second-order density matrix, that appears in the equation of motion for the first-order one, is computed via semiclassical dynamics. Here we showed that running semiclassical dynamics on the whole system does approximately capture double-excitations. Future work includes retaining exact dynamics for the one-body terms, leaving semiclassics just for the correlation component, according to the original prescription, to see if the accuracy of the states of double-excitation character are improved. 

In conclusion, there are many fascinating things to be learnt and discovered about states of double-excitation character! We hope that the findings here, and in the earlier work reviewed in the introduction, will spur more interesting investigations, with both practical and fundamental consequences.

We acknowledge support from the National Science Foundation
(CHE-0647913), the Cottrell Scholar Program of Research Corporation
and the NIH-funded MARC program (SG and CC).

\end{document}